# Logic Compatible High-Performance Ferroelectric Transistor Memory

Sourav Dutta[1]*, Huacheng Ye[1], Abhishek Khanna[1], Yuan-Chun Luo[2], Lillian Pentecost[3], Akif A. Khandker[1], Wriddhi Chakraborty[1], Gu-Yeon Wei[3], David Brooks[3], Michael Niemier[5], Xiaobo Sharon Hu[5], Shimeng Yu[2], Kai Ni[4] and Suman Datta[1]*

[1]Department of Electrical Engineering, University of Notre Dame, Notre Dame, IN 46556, USA

[2]School of Electrical and Computer Engineering, Georgia Institute of Technology, Atlanta, GA 30332, USA.

[3]School of Engineering and Applied Sciences, Harvard University, Cambridge, MA 02138

[4]Department of Electrical and Microelectronic Engineering, Rochester Institute of Technology, Rochester, NY 14623 USA

[5]Department of Computer Science and Engineering, University of Notre Dame, Notre Dame, IN 46556, USA

* Corresponding author: sdutta4@nd.edu, sdatta@nd.edu

# ABSTRACT

Silicon ferroelectric field-effect transistors (FeFETs) with low-k interfacial layer (IL) between ferroelectric gate stack and silicon channel suffers from high write voltage, limited write endurance and large read-after-write latency due to early IL breakdown and charge trapping and detrapping at the interface. We demonstrate low voltage, high speed memory operation with high write endurance using an IL-free back-end-of-line (BEOL) compatible FeFET. We fabricate IL-free FeFETs with 28nm channel length and 126nm width under a thermal budget <$400^0$C by integrating 5nm thick $Hf_{0.5}Zr_{0.5}O_2$ gate stack with amorphous Indium Tungsten Oxide (IWO) semiconductor channel. We report 1.2V memory window and read current window of $10^5$ for program and erase, write latency of 20ns with $\pm$2V write pulses, read-after-write latency <200ns, write endurance cycles exceeding $5x10^{10}$ and 2-bit/cell programming capability. Array-level analysis establishes IL-free BEOL FeFET as a promising candidate for logic-compatible high-performance on-chip buffer memory and multi-bit weight cell for compute-in-memory accelerators.

**INTRODUCTION**

The paradigm of computing is constantly evolving to enable the vision of edge artificial intelligence (e-AI) where trillions of connected smart devices are pervasively being integrated into our daily life. These devices will constantly measure physical parameters and process them locally to provide actionable intelligence in real-time [1]. Data-centric local computing at the point of data acquisition can minimize bandwidth and latency issues associated with communication and enable critical applications that require sensing, processing and feedback in rapid succession. Today's primary candidate for on-chip memory - static random-access memory (SRAM) is volatile, suffers from large footprint area (120-150F$^2$) and high standby leakage power. The embedded dynamic random-access memory (eDRAM) is also volatile and requires frequent refresh, thus incurring high power dissipation. Even with the continued advancement in transistor scaling, leading to improvement in processor speed and energy-efficiency, the performance of the overall system is limited by the extraordinary volume of data traffic between the processor and the off-chip memory. New architectural innovations such as compute-in-memory (CIM) promises an advantage in energy and latency by moving compute closer to the data residing in the memory block. The holy grail of CIM is high bandwidth, low-latency access to on-chip embedded non-volatile memory (eNVM) that is programmable with logic-compatible voltages, has the footprint of a single transistor, exhibits infinite endurance and is BEOL compatible for integrating on top of high-performance advanced logic to realize the true potential of M3D-CIM[2], [3].

Harnessing the polar orthorhombic phase Pca2$_1$ in thin films of doped Hafnium Oxide (HfO$_2$), one can construct a one transistor non-volatile ferroelectric memory (FeFET) that can serve as a logic-compatible, high-performance embedded non-volatile memory [2]. Recent demonstrations of robust memory operation in FeFETs with high density integration in relatively scaled CMOS technology nodes (28nm planar bulk [4] and 22nm FD-SOI [5]) has provided legitimacy of FeFET as a contender for eNVM with best-in-class energy efficiency, latency and footprint area for CIM applications standing next to spin-transfer-torque magnetic random-access memory (STT-MRAM), resistive random-access memory (RRAM) and phase change memory (PCM) [2], [6]–[8]. However, FeFETs still face several key roadblocks such as logic-compatible programming, write endurance, read-after-write latency and BEOL-compatible low temperature processing [9]–[13] for integration in the BEOL.

Fig. 1(a) shows the schematic of a silicon FeFET with an undesired interfacial layer (IL) with low-dielectric constant (k=3.9 for $SiO_2$) between the silicon channel and ferroelectric HZO gate stack. The limited write endurance in the conventional FeFET predominantly arises due to the presence of the low-k IL. As elucidated in the energy band diagram (see supplementary section **S1** for details) in Fig. 1(c, i), during the write operation of a conventional FeFET, most of the programming voltage in the gate stack drops across the IL which may reach as high as $E_{IL}$=10MV/cm [2], [14]. The high electric-field stresses the IL along with continuous charge trapping and trap generation at the either the interface between the IL and silicon or the FE-IL interface that tend to limit the write endurance in the range of $10^4$-$10^6$ [9]–[12]. It is to be noted that, the ferroelectric HZO in a stand-alone metal-ferroelectric-metal (MFM) capacitor withstands a write endurance of over $10^{10}$ cycles (see supplementary section **S2** for endurance measurements on MFM capacitors) [14], [15]. Thus, improving the write endurance of FeFET requires us to either re-explore the known techniques of IL engineering in HKMG technology such as reducing the IL thickness through scavenging [16] and increasing the dielectric constant of IL [17], or re-imagine novel ways to build an IL-free FeFET with high write endurance. The latter can be achieved by fabricating a back-gated, channel-last FeFET where an oxide semiconductor channel is grown directly on top of the ferroelectric HZO gate oxide, analogous to the fabrication of thin-film transistors (TFTs) used commercially in flat-panel display application [8], [18]–[22]. Fig. 1(b) shows a schematic of an IL-free FeFET along with the calculated band diagram in Fig. 1(c, ii). The absence of IL allows lowering the write voltage to logic-compatible voltage levels and significantly reduces the charge trapping, and trap generation at the interface between FE and channel, thereby improving the write endurance.

The absence of IL also reduces the read-after-write latency of the FeFET which is predominantly due to electron and hole trapping at the interface between FE-IL and IL-silicon, respectively. Prior work on conventional FeFET have reported a large read-after-write latency due to charge trapping at the interface of HZO/IL and IL/silicon during the write operation associated with ferroelectric switching. During programming with positive gate voltage, electrons get trapped primarily at the HZO/IL interface. As shown in the energy band diagram in Fig. 1(c, iii), during the read operation, a V-shaped energy potential is formed at the HZO/IL interface. The large read-after-write latency is set by the time required for the electrons to de-trap via tunneling through the IL which has been measured to be in the range of 10ms. Similar scenario also arises for the erase operation where the negative gate voltage favors the hole trapping primarily at the interface of IL and silicon channel, requiring close to $100\mu s$ time delay to de-trap [9], [14]. On the other hand, as shown in the band diagram in Fig. 1(c, iv), the absence of IL allows fast de-trapping, resulting in low read-after-write latency.

Here, we present a novel IL-free BEOL compatible back-gated, channel last FeFET with ultra-scaled dimensions of Lg=28nm and W=126nm. The FeFET is fabricated by depositing a 5nm thick amorphous Tungsten-doped Indium Oxide (IWO) semiconductor channel material directly on top of 5nm ferroelectric Zirconium-doped $HfO_2$ ($Hf_{0.5}Zr_{0.5}O_2$) gate oxide, thus avoiding the creation of the undesired low-k IL. Prior to IWO thin film deposition, the $Hf_{0.5}Zr_{0.5}O_2$ film undergoes a crystallization anneal at $400^0$C for 300s using a sacrificial capping layer (SCL) to stabilize the polar orthorhombic phase (see supplementary section **S2** for details). The entire fabrication is done under a low thermal budget of $400^0$C. The details are described in the **Methods** section. We choose IWO as the n-type amorphous metal oxide semiconductor (AOS) channel material since it provides a higher field-effect electron-mobility of 20 $cm^2$/V-s compared to other contemporary AOS materials such as IGZO (~ 10 $cm^2$/V-s) under low thermal budget processing. Furthermore, IWO has a higher threshold voltage ($V_T$) stability compared to IGZO due to higher oxygen-bond dissociation energy of Tungsten that acts as both stabilizer and electron donor [3], [23], [24].

**RESULTS**

Fig. 1(d) shows a schematic of the fabricated BEOL FeFET consisting of a Tungsten (W) local back-gate, 5nm HZO as the ferroelectric gate oxide, 5nm IWO as the semiconductor channel and Palladium (Pd) as the source and drain contacts. Fig. 1(e) shows the top view false-colored SEM image of the fabricated device and a zoomed-in image highlighting the ultra-scaled device dimensions of Lg = 28nm and width = 126nm. Fig. 1(f) shows a cross-sectional TEM of the FeFET, confirming the ultra-scaled channel length and thickness of individual layers. A high-resolution (HR) TEM cross-section image of the gate-source overlap region shows the distinct amorphous nature of the IWO channel material as opposed to the crystalline nature of the HZO layer. The elemental composition is further confirmed from the EDS elemental mapping of the fabricated device as shown in Fig. 1(g).

The BEOL FeFET operates as a junction-less FET with the gate-to-drain and gate-to-source overlap regions playing a crucial role in the memory operation of the device. As shown in the HRTEM image in Fig. 1(f), the overlap regions resemble a metal-ferroelectric-semiconductor-metal (MFSM) structure where most of the applied electric field during the program and erase operation is concentrated and induces effective polarization switching in the HZO layer. In addition, there exists fringing electric fields inside the channel region away from the source and drain contacts that result in polarization switching in the HZO layer underneath the channel. This implies that that the read voltage memory window will monotonically increase with decreasing channel length in such a back-gated FeFET device geometry [18]. During the programming state, the polarization charge in HZO can accumulate a large channel charge in IWO, thus turning on the transistor. In the erase state, the channel charge in IWO is depleted by the polarization charge in HZO, thus turning off the transistor.

Unlike perovskite-based ferroelectrics, the polar orthorhombic phase is retained in ultra-thin layers of doped $HfO_2$ up to 1 nm thickness [25] (see supplementary section **S2** for characterization of 5nm thick MFM capacitors). This allows lowering of the programming voltage of HZO-based FeFETs from $\pm 4V$ for 10nm thickness [26] to $\pm 3.3V$ for 4.5nm [12]. However, in Si FeFETs, most of the programming voltage in the gate stack drops across the low-k IL between the HZO and silicon. On the other hand, an IL-free FeFET enables further lowering of the programming voltage for the same ferroelectric thickness. Fig. 2(a) shows the dual-sweep DC Id-Vg characteristic of the BEOL FeFET with W/L = 126nm/28nm and $V_{DS}$ = 0.2V. A memory window (MW) over 1.2V is obtained with a sweep of $\pm$2V. We report a read current $I_{LVT}$ = 2$\mu$A for the program (or low $V_T$) state at read voltage $V_{read}$ = -0.2V and a read current $I_{HVT} \sim$ 10pA for the erase (or high $V_T$) state which provides a high read current window of $I_{LVT}/I_{HVT} > 10^5$. Fig. 2(b) shows the measured subthreshold slope (SS) for the dual sweep. In the forward sweep, we measure a SS $\sim$ 95mV/dec while a steep slope of SS $\sim$ 4mV/dec is measured is the reverse sweep owing to the ferroelectric switching kinetics [27]. Such a steep-slope, also known as transient negative capacitance (NC) effect, arises due to the difference between the rate of change of polarization switching and rate of change of screening or compensating free charge [28]–[30].

Next, we characterize the fast read and write operation of the BEOL FeFET memory. In most studies, the memory characteristics of a FeFET is determined in terms of threshold voltage ($V_T$) and MW measurement. In real applications such as compute-in-memory, the actual current being read out of the memory device (after program and erase) will play the critical role of determining the overall array-level efficiency such as energy and latency. Hence, here we adopt a one-shot fast read-out scheme of BEOL FeFET by applying a short read pulse to the gate of the FeFET following the write (program or erase) pulse. For either program or erase, we apply write pulses of amplitude $\pm$2V and pulse-width of 50ns at the gate of the FeFET. During the write operation, both the source and drain of the FeFET are grounded by applying $V_{DS}$ = 0V. Following the write operation, we apply $V_{read}$ = -0.2V to the gate of FeFET while a $V_{DS}$ = 0.2V is applied across the source and drain terminals to measure the drain current $I_D$. Figs. 2(c, d) show the time-domain response of the FeFET during the program and erase operation. For the programming scenario, we see a sharp increase in read current $I_D$ over 1$\mu$A during the read-out, whereas, for the erase case, the read current $I_D$ remains close to 1nA. Note that a key characteristic of the IL-free BEOL FeFET is the ability to read the memory state almost instantaneously after the write operation without the need for the charge trapped at the FE-IL interface during write phase to detrap before read operation can commence. In contrast to the conventional Si FeFET that in some cases require as high as almost 10ms of read-after-write latency [9], [14], we demonstrate fast read-after-write operation with a latency of 200ns (limited by the pulsed I-V measurement setup). We also measure the cycle-to-cycle variation by performing repeated program and erase operations with $\pm 2$V and 50ns pulses. Fig. 2(e) shows the distribution of the measured $I_D$ for 140 repeated cycles of program and erase, showing a clear separation of memory states. We quantified the switching speed of the BEOL FeFET in terms of the write pulse amplitude and pulse width required to obtain a $I_{LVT} > 1\mu A$, $I_{HVT} < 10$nA and $\Delta I_D > 1\mu A$. Fig. 2(f) shows the exponential dependence of the required pulse width on the applied voltage. We obtain symmetrical switching behavior for both program and erase operation with write latency as small as 20ns for $\pm 2$V.

Next, we investigate the write endurance characteristics of the BEOL-compatible IL-free FeFET. We use fast bipolar write pulses of $\pm 2$V and 20ns for endurance cycling with periodic reading of the memory state ($I_{LVT}$ and $I_{HVT}$ at $V_{read}$ for program and erase) and DC Id-Vg measurement to monitor the device characteristics. Fig. 3(a) shows the readout current for program and erase states as a function of endurance cycles. Post $5 \times 10^{10}$ write cycles, the FeFET maintains an $I_{LVT} > 1\mu A$, while the $I_{HVT}$ gradually increases from 10pA up to 100nA. Still, after $5 \times 10^{10}$ cycles, the device maintains a clear separation of the read current level for the program and erase states with a read current window of $I_{LVT}/I_{HVT} > 10$ as seen in Fig. 3(b). Fig. 3(c) shows the DC Id-Vg characteristics of the device measured periodically post endurance cycling. In contrast to rapid degradation and permanent failure exhibited by Si FeFETs after $10^4$-$10^6$ cycles [9]–[13] due to catastrophic breakdown of the IL, we see a graceful degradation in the case of the IL-free IWO FeFET with increasing subthreshold slope and a shift in the threshold voltage $V_T$. This can be correlated to an increase in the generation of interface trap density ($N_{it}$) at the HZO/IWO interface (see supplementary section **S4** for details). The anti-clockwise hysteresis remains preserved in the FeFET, exhibiting robust ferroelectricity in the 5nm HZO gate oxide with no sudden breakdown observed whatsoever. We also characterized the read endurance of the BEOL FeFET. After programming (or erasing) the device with a positive (or negative) write pulse, we apply continuous read pulses with $V_{read} \sim$ -0.2V. We witness no read disturb and degradation in $I_{LVT}$ and $I_{HVT}$ as seen in Fig. 3(d). Fig. 3(e) shows the excellent memory retention characteristic of the BEOL FeFET, measured at $25^0$C for $10^3$s.

Finally, we demonstrate the multi-bit per cell programming characteristic of the FeFET that allow storage of multi-bit weight matrix on-chip for performing fast and energy-efficient vector-matrix multiplication *in situ* to accelerate deep neural network (DNN) inference workloads. We utilize the voltage-driven partial polarization switching in the ferroelectric HZO layer to implement the multi-bit programming capability. As shown in Fig. 4, we use the amplitude modulation scheme. Following an erase operation with -2V, programming pulses with increasing amplitudes of 1V, 1.3V, 1.7V and 2V are applied to gradually modulate the conductance state of the device and program it to four distinct states representing bits '00', '01', '10' and '11', respectively. Fig. 4 shows the four tightly distributed current levels measured over 50 programming cycles, clearly exhibiting a 2bit/cell memory characteristic.

**DISCUSSION**

The back-gated, channel last process for fabricating FeFETs provides an effective pathway to constructing an IL-free BEOL compatible logic-compatible, high-performance ferroelectric memory that mitigates several challenges faced by conventional Si FeFETs. In Table 1(a), we enumerate a comprehensive device-level comparison between the state-of-the-art results reported on FEOL silicon FeFETs [5], [17], the IL-free BEOL FeFET and other contemporary BEOL FeFET embodiments [19]–[21]. We report the best-in-class device performance with a low operating voltage of $\pm$2V, high-speed read and write operation and endurance cycling exceeding $5\times10^{10}$ write cycles in an extremely scaled device geometry of W*L=$0.0035\mu m^2$ with a ferroelectric gate oxide of 5nm thickness. Additional thinning of the ferroelectric gate dielectric to 3nm will further reduce the programming voltage down to 1.5V. This makes the IL-free BEOL FeFET a promising CMOS logic compatible high-performance on-chip buffer memory and multi-bit weight cell for future compute-in-memory accelerators.

We perform system-level benchmarking to compare the performance of various eNVMs and SRAM for storing multi-bit weight matrix on-chip to realize key operations such as vector-matrix-multiplication during inference. We use an 8-bit inference workload using the VGG-8 model on CIFAR-10 dataset. We use the DNN+NeuroSim framework [31] which is a widely used to benchmark and evaluate CIM accelerators (see supplementary section **S5** for details). We use a cell structure consisting of 1T-1BEOL FeFET in a pseudo-crossbar array [8]. Table 1(b) shows the comparison between  IL-free BEOL FeFET, FEOL FeFET [6], RRAM [32], STT-MRAM [33] and SRAM. We report the best-in-class energy efficiency of 255.3 TOPs/W in the densest footprint area of 119.9mm$^2$ for a single CIM array which provides a 10x improvement in energy-efficiency and 4x improvement in area compared to SRAM.

In this benchmarking exercise, we assume a 128KB SRAM-based on-chip global buffer memory (scratchpad memory) which can fit the maximum size of activation for VGG-8 on CIFAR-10 dataset with a single data transfer from off-chip DRAM. Larger datasets such as ImageNet would require a much denser on-chip global buffer memory. Furthermore, in the benchmarking, the weights are assumed to be stored on-chip in the synaptic array prior to the inference task. For lightweight smart edge devices with area-constraint design, this may not hold and only a part of the weight matrix can be stored on-chip [34]. For example, more powerful DNN models for image processing such as ResNet [35] have around 12-30 MB of parameters that would require either frequent, costly data movement from off-chip DRAM to load weights per inference or building denser on-chip global buffer memories. We show that the IL-free FeFET can serve as an incredibly dense on-chip storage for parameters and thus enable larger DNN models to be efficiently executed on the CIM accelerator without fetching DNN weights from the DRAM during inference. This in turn can further maximize inference efficiency [36]. Table 1(c) shows the performance comparison between SRAM, STT-MRAM [37] and IL-free BEOL FeFET when used as a 32MB global buffer memory to support a ResNet50 DNN model. The array-level characteristics are computed using NVSim framework [38]. Compared to SRAM, IL-free FeFET increases the storage density by over 6x with similar read/write access energy and latency. IL-free FeFET also significantly improves the memory density over iso-capacity when compared to STT-MRAM, a prime candidate for scratchpad and last-level-cache memories.

**METHODS**

**Device Fabrication.** We fabricated FeFETs starting with a patterned W local back gate formation. We use plasma enhanced atomic layer deposition (PEALD) to deposit 5nm thick HZO at a deposition temperature of $250^0$C. For strain-induced stabilization of ferroelectric orthorhombic phase in HZO, we sputter-deposit W as a sacrificial capping layer (SCL). We perform rapid thermal anneal (RTA) at $400^0$C for 300s in $N_2$. Subsequently, W SCL is removed used etching. Next, 1% by weight W-doped amorphous $In_2O_3$ (IWO) channel of 5nm thickness was deposited using RF magnetron sputtering in the presence of 0.02Pa excess $O_2$ at room temperature. This is followed by patterning and liftoff of the active IWO channel. Next, 30 nm thick Pd was patterned as source and drain electrodes, followed by a $150^0$C $N_2$ anneal for contact resistance improvement. The channel conductivity significantly depends on the thickness of IWO and $O_2$ partial pressure used during deposition of IWO. As such, we also fabricated non-ferroelectric IWO transistors with 5nm $HfO_2$ gate stack for comparison. Supplementary section **S3** shows the measured transfer characteristics and dependence of the threshold voltage ($V_T$) of the baseline transistors fabricated with different IWO channel thickness and different $O_2$ partial pressures.

**Device Characterization.** All electrical characterizations are performed using Keithley 4200-SCS probe station, Agilent 81150A pulse generator and Agilent DSO9104A oscilloscope.

**AUTHOR CONTRIBUTIONS**

S. Dutta and S. Datta conceived the idea. S. Dutta, A. Khanna., A. Khandker and W. C. performed the experiments. H. Y. performed the fabrication. B. R. supervised the ALD process. K. N. performed the TCAD simulations. Y. L. and S. Y. performed the benchmarking for inference workload using DNN+Neurosim. L. P. G. W. and D. B. performed the benchmarking on global buffer memory using NVSim framework. M. N and X. S. H. participated in useful discussion. S. Dutta and S. Datta wrote the manuscript.

**ACKNOWLEDGEMENT**

This work was supported by ASCENT, one of six centers in JUMP, sponsored by DARPA and the Semiconductor Research Corporation (SRC), and IMPACT center in nCORE, sponsored by SRC.

**Figures**

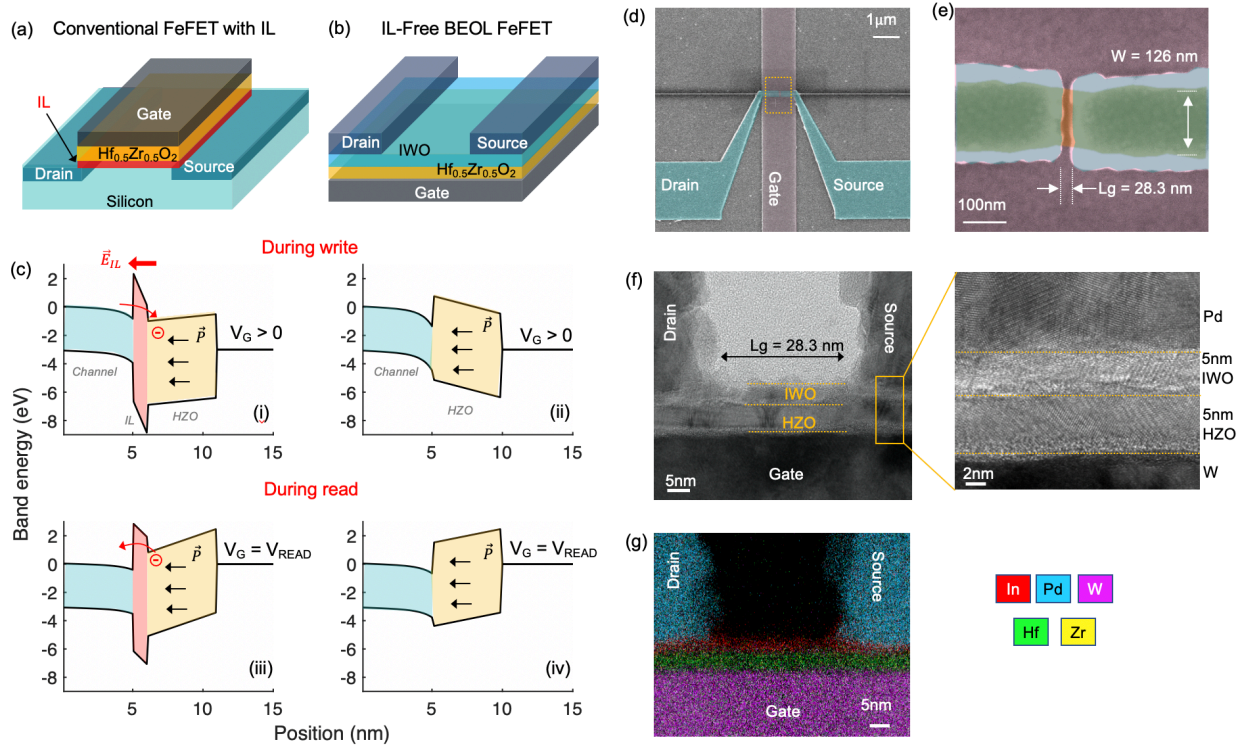

**Figure 1**. IL-free BEOL FeFET. (a) Schematic of a conventional FeFET with a low-k interfacial layer (IL) and (b) IL-free FeFET. (c) Calculated energy band-diagrams for the two structures during write and read operations. The high electric-field ($E_{IL}$) stress across the IL along with continuous charge trapping and trap generation at the interface tend to limit the write endurance of conventional FeFET. The presence of IL also sets a large charge de-trapping time constant that increases the read latency in conventional FeFET. The absence of IL allows lowering the write voltage and significantly reduces the interfacial charge trapping, thereby improving the write endurance and read latency. (d) Schematic of the fabricated device consisting of a local Tungsten (W) back-gate, 5nm $Hf_{0.5}Zr_{0.5}O_2$ as the ferroelectric gate dielectric, 5nm Tungsten (W)-doped amorphous $In_2O_3$ (IWO) as the semiconductor channel and Palladium (Pd) as the source and drain contacts. (e) Top-view false colored SEM image of the fabricated device showing the ultra-scaled

device dimensions of Lg = 28.3nm and width = 126nm. (f) Cross-sectional TEM image confirms the ultra-scaled channel length and thickness of individual layers. High-resolution TEM image of the gate-to-source overlap region shows the distinct amorphous nature of the IWO channel material as opposed to the crystalline nature of the HZO layer. (g) Elemental composition of the fabricated device shown in the EDS elemental mapping image.

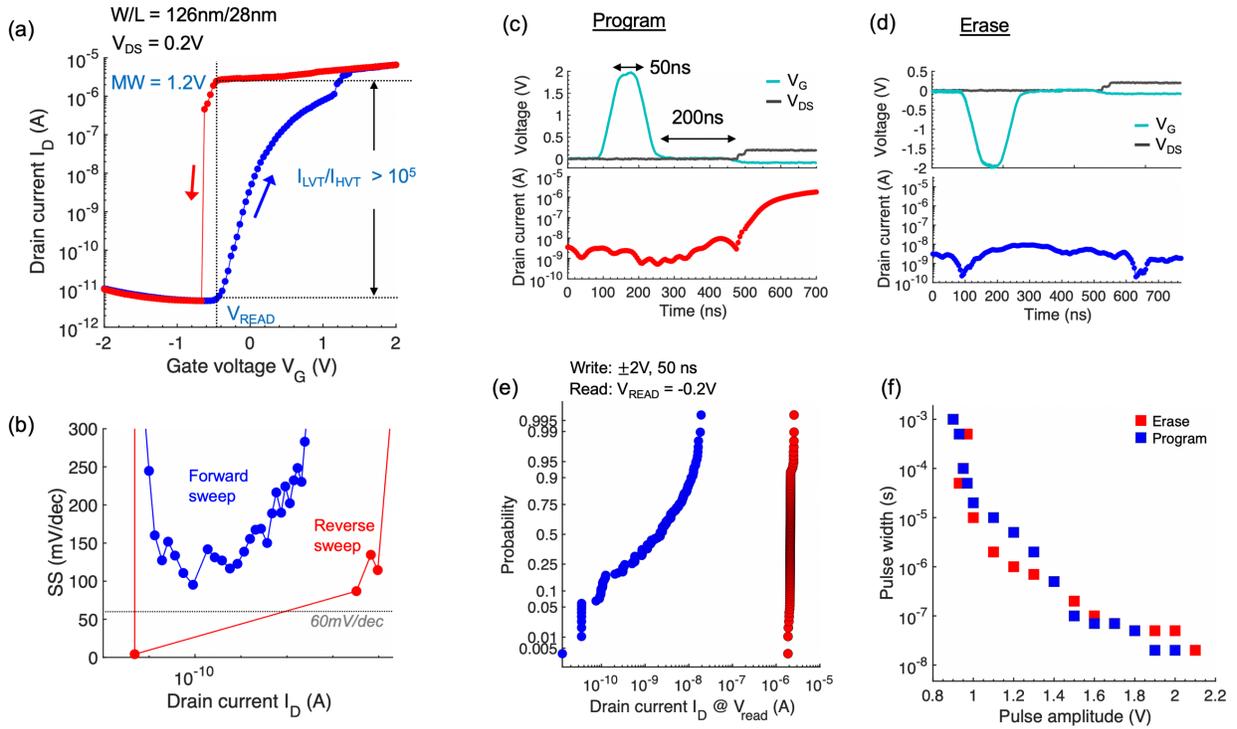

**Figure 2**. Memory characterization of the ultra-scaled BEOL FeFET with W/L=126nm/28nm. (a) Dual sweep DC Id-Vg characterization with $\pm$2V shows a memory window (MW) of 1.2V. The device exhibits a high read current $I_{LVT} = 2\mu A$ for the program state and a low read current $I_{HVT} \sim$ 10pA for the erase state with read current window of $I_{LVT}/I_{HVT} > 10^5$. (b) Measured subthreshold slope (SS) with an SS=95mV/dec for the forward sweep and a steep slope of SS=4mV/dec for the reverse sweep. (c, d) Time-domain response of the device during program and erase operation. A write voltage of $\pm$2V and 50ns is applied to the gate of FeFET to program or erase, followed by a fast read out using a $V_{read} = -0.2V$ at the gate and $V_{DS} = 0.2V$ with a 200ns read-after-write latency. (e) Distribution of the measured $I_D$ for 140 repeated program and erase cycles, showing a clear separation of memory states upon cycling. (f) Switching speed, defined as the write pulse amplitude and pulse width required to obtain a $I_{LVT} > 1\mu A$, $I_{HVT} < 10nA$ and read current window

$\Delta I_D > 1\mu A$. An exponential dependence of the required pulse width on the applied voltage is seen

write latency as small as 20ns for $\pm 2V$.

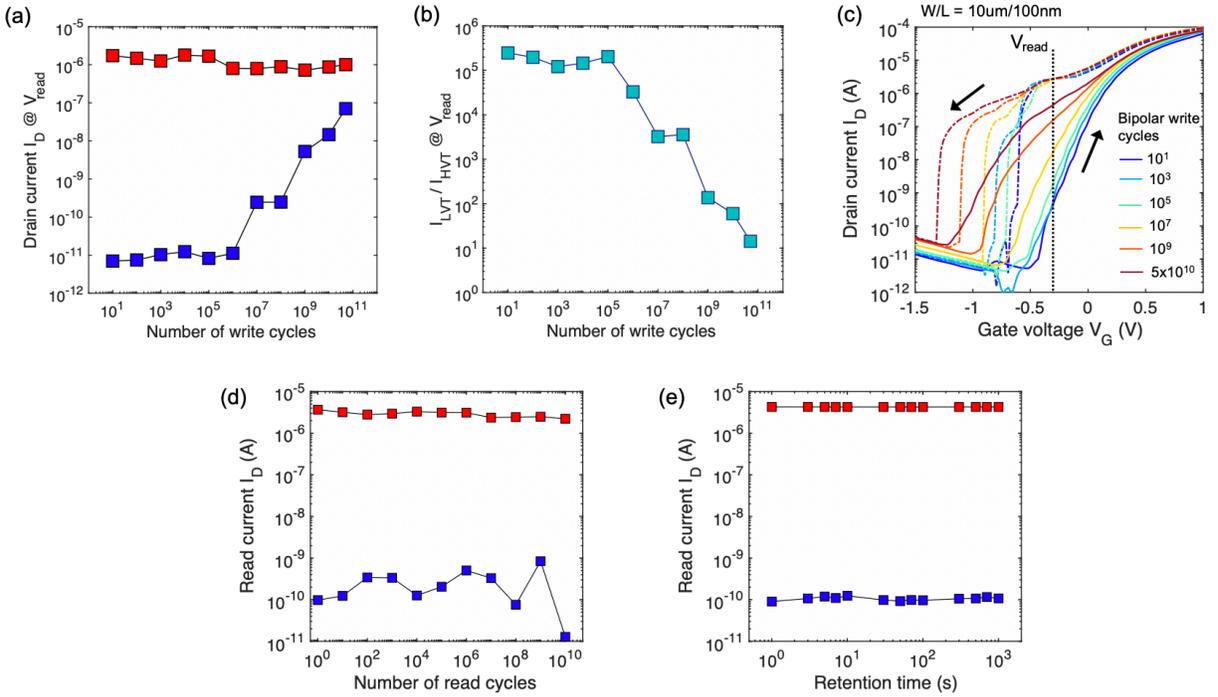

**Figure 3**. Endurance characterization of the BEOL FeFET. (a) Bipolar stress of $\pm 2V$ and 20ns is applied to the device. The measured drain current for program and erase states as a function of endurance cycles shows that even after $5 \times 10^{10}$ write cycles, the device maintains a $I_{LVT} > 1 \mu A$ while the $I_{HVT}$ slowly increases from 10pA up to 100nA. (b) Measured read current window of the device with endurance cycling. Even after $5 \times 10^{10}$ cycles, the device maintains a clear separation of current level for the program and erase states with $I_{LVT}/I_{HVT} > 10$. (c) DC Id-Vg characteristics of the device measured periodically post endurance cycling. A graceful degradation with increasing subthreshold slope and a shift in the threshold voltage $V_T$ is seen. However, the anti-clockwise hysteresis is retained in the BEOL FeFET, exhibiting robust ferroelectric switching in the 5nm HZO gate oxide with no sudden breakdown and permanent device failure. (d) Read endurance characterization of the BEOL FeFET. (a) After programming (or erasing) the device

with a positive (or negative) write pulse, continuous read pulses with $V_{read} \sim -0.2V$ are applied to readout the memory state. No read disturb and degradation in $I_{LVT}$ and $I_{HVT}$ is seen. (e) Measured memory retention of the BEOL FeFET at $25^0C$, showing excellent retention characteristics with no degradation.

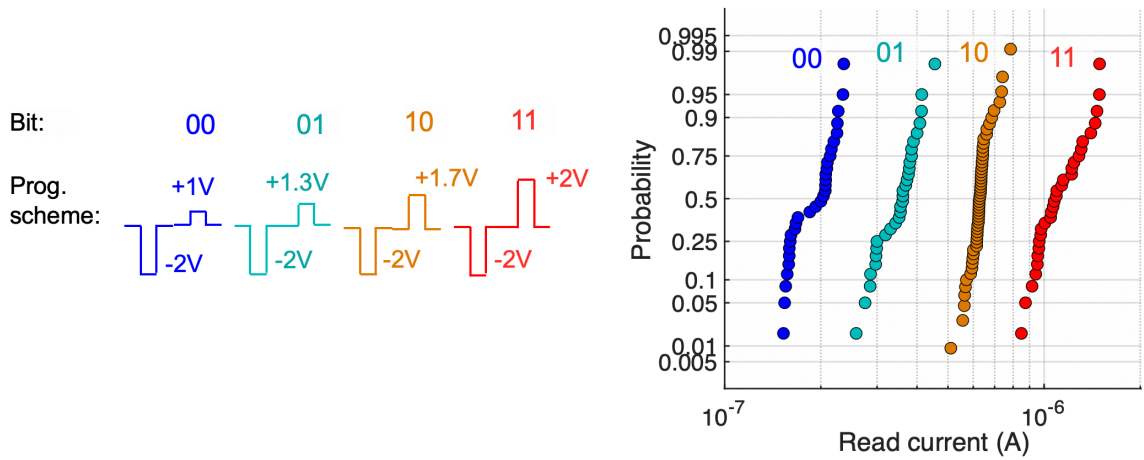

**Figure 4**. Multi-bit per cell characteristics of the FeFET that will allow storing multi-bit weight matrix on-chip. We use the amplitude modulation scheme to induce voltage-driven partial polarization switching in the ferroelectric HZO layer and implement multi-bit programming capability. Following an erase operation with -2V, programming pulses with increasing amplitudes of 1V, 1.3V, 1.7V and 2V are applied to gradually modulate the conductance state of the device and program it to four distinct states representing bits '00', '01', '10' and '11', respectively. A 2bit/cell memory characteristic with four tightly distributed current levels is clearly seen for repeated programming of the device over 50 times.

| (a) | Device Level Benchmarking | | | |
|---|---|---|---|---|
| | Conventional FeFET with IL | IL-Free BEOL FeFET | | |
| | [5, 16] | [18, 19] | [20] | This Work |
| Channel Material | Silicon | IGZO | $WO_x$ | IWO |
| Ferroelectric | 4.5nm HZO | 15nm HZO | 10nm HZO | 5nm HZO |
| Maximum process temperature | $500^0C$ | $500^0C$ | $375^0C$ | $400^0C$ |
| Scaling (W/L) | 80nm/20nm | $30\mu m/10\mu m$ | $20\mu m/5\mu m$ | 126nm/ 28nm |
| Subthreshold slope (SS) | 100mV/dec | 90 mV/dec | / | 95mV/dec |
| Memory Window | 1.5V | 0.5V | 1V | 1.2V |
| $I_{LVT}/I_{HVT}$ ratio | $10^5$ | $10^5$ | / | $>10^5$ |
| Programming voltage | $\pm3V$ | $\pm3V$ | $\pm3.6V$ | $\pm2V$ |
| Write latency | 250ns | 500 us | 5us | 20ns |
| Write endurance | $>10^{10}$ | $>10^8$ | / | $>10^{10}$ |
| Read latency | 80us | / | / | 200ns |
| Retention | $>10^4$ s | $>4x10^3$s | $>10^4$ s | $>10^3$ |
| Multi-bit/Cell | 1 | 1 | 4 | 2 |

| (b) | Inference Benchmarking with VGG-8 model on CIFAR-10 | | | | |
|---|---|---|---|---|---|
| | SRAM | RRAM | STT-MRAM | Conventional FeFET | IL-Free BEOL FeFET |
| Bit/Cell | 1 | 2 | 1 | 2 | 2 |
| Cell area (F2) | 600 | 60 | 100 | 26 | 15 |
| Single CIM array area (mm2) | 481.69 | 475.79 | 792.99 | 206.18 | 118.95 |
| L-by-L Dynamic Energy (uJ) | 46.28 | 21.94 | 79.98 | 7.67 | 4.74 |
| L-by-L latency (ms) | 1.3 | 0.76 | 1.31 | 0.7 | 0.67 |
| Energy Efficiency (TOPs/W) | 25.4 | 55.90 | 15.34 | 158.83 | 255.33 |
| Throughput (TOPs) | 0.95 | 1.61 | 0.94 | 1.76 | 1.83 |

| (c) | Benchmarking 32MB Global Buffer Memory for ResNet50 | | |
|---|---|---|---|
| | SRAM | STT-MRAM | IL-Free BEOL FeFET |
| Global buffer area (mm$^2$) | 21.68 | 9.92 | 3.09 |
| Memory density (MB/mm$^2$) | 1.48 | 3.23 | 10.3 |
| Write energy (pJ) | 8.2 | 55.8 | 39.1 |
| Write latency (ns) | 13 | 5.9 | 21.7 |
| Read energy (pJ) | 4.9 | 29.0 | 9.0 |
| Read latency (ns) | 13 | 5.8 | 21.3 |

**Table 1**. Comprehensive benchmarking of the IL-free BEOL FeFET. (a) Device-level comparison of the IL-free BEOL FeFET demonstrated in the work with the state-of-the-art results reported on FEOL silicon FeFETs and other contemporary BEOL FeFET proposals. Overall, BEOL FeFET exhibits the best-in-class device performance with a low operating voltage, high-speed memory operation and high write endurance in a highly scaled device geometry. (b) System-level benchmarking using the DNN+NeuroSim framework [31] to compare the performance of various eNVMs and SRAM for storing multi-bit weight matrix on-chip. We use an 8-bit inference

workload using the VGG-8 model on CIFAR-10 dataset. BEOL FeFET exhibits the best-in-class energy efficiency in the densest footprint area with 10x improvement in energy-efficiency and 4x improvement in area compared to SRAM. (c) Array-level performance comparison performed using NVSim framework [37] for a 32MB global buffer memory used to support a ResNet50 DNN model. Compared to SRAM, IL-free FeFET increases the storage density by over 6x with similar read/write access energy and latency.